\documentclass[prb,showpacs,superscriptaddress,floatfix,twocolumn]{revtex4}
\usepackage{amsfonts}
\usepackage{stmaryrd}
\usepackage{bbm}
\usepackage{mathrsfs}
\usepackage{tipa}
\usepackage{amssymb}
\usepackage{txfonts}
\usepackage{graphicx}
\usepackage{dcolumn}
\usepackage{epstopdf}
\usepackage{array}

\newcommand{\PreserveBackslash}[1]{\let\temp=\\#1\let\\=\temp}
\newcolumntype{C}[1]{>{\PreserveBackslash\centering}p{#1}}
\newcolumntype{R}[1]{>{\PreserveBackslash\raggedleft}p{#1}}
\newcolumntype{L}[1]{>{\PreserveBackslash\raggedright}p{#1}}

\begin{document}

\newcommand*{\cm}{cm$^{-1}$\,}


\title{Optical spectroscopy study of Nd(O,F)BiS2 single crystals}
\author{X. B. Wang}
\author{S. M. Nie}
\author{H. P. Wang}
\author{P. Zheng}
\author{P. Wang}
\author{T. Dong}
\affiliation{Beijing National Laboratory for Condensed Matter
Physics, Institute of Physics, Chinese Academy of Sciences,
Beijing 100190, China}

\author{H. M. Weng}
\affiliation{Beijing National
Laboratory for Condensed Matter Physics, Institute of Physics,
Chinese Academy of Sciences, Beijing 100190, China}
\affiliation{Collaborative Innovation Center of Quantum Matter, Beijing, China}

\author{N. L. Wang}
\affiliation{Beijing National
Laboratory for Condensed Matter Physics, Institute of Physics,
Chinese Academy of Sciences, Beijing 100190, China}
\affiliation{Collaborative Innovation Center of Quantum Matter, Beijing, China}

%

\begin{abstract}

We present an optical spectroscopy study on F-substituted NdOBiS$_2$ superconducting single crystals grown using KCl/LiCl flux method. The measurement reveals a simple metallic response with a relatively low screened plasma edge near 5000 \cm. The plasma frequency is estimated to be 2.1 eV, which is much smaller than the value expected from the first-principles calculations for an electron doping level of x=0.5, but very close to the value based on a doping level of 7$\%$ of itinerant electrons per Bi site as determined by ARPES experiment. The energy scales of the interband transitions are also well reproduced by the first-principles calculations. The results suggest an absence of correlation effect in the compound, which essentially rules out the exotic pairing mechanism for superconductivity or scenario based on the strong electronic correlation effect. The study also reveals that the system is far from a CDW instability as being widely discussed for a doping level of x=0.5.

\end{abstract}


\maketitle

\section{Introduction}
The newly discovered bismuth oxysulfide superconductors have attracted much attention in the past two years. Soon after the discovery of superconductivity at 4.4 K in Bi$_4$O$_4$S$_3$,\cite{E1,E6} several new members of this family, e.g. ReO$_{1-x}$F$_x$BiS$_2$ (Re= La, Ce, Pr, Nd, Yb), Sr$_{1-x}$La$_x$FBiS$_2$, and La$_{1-x}$M$_x$OBiS$_2$ (M= Ti, Zr, Hf, Th), were found.\cite{E2,E4,E13,E15,E16,E19,E21,E36} Those compounds have layered structures consisting of alternate stacking of superconducting BiS$_2$ layers and different blocking layers. The stacking structure is analogous to those of high-T$_c$ cuprates \cite{cuprates} and Fe-based superconductors \cite{iron-based}. Typically, superconductivity arises when charge carriers are introduced into the parent compound, and the superconducting transition temperature T$_c$ reaches a maximum value at the nominal doping level of x=0.5.\cite{E2}

First principle calculations indicate that the parent compound is a semiconductor with an energy gap of $\sim$0.8 eV, and superconductivity induced by electron doping is derived mainly from the 6p$_x$/p$_y$ orbitals in the BiS$_2$ layers.\cite{T1,T3} At the doping level of x=0.5, a good nesting between the large parallel Fermi-surface segments with wave vector $\textbf{q}=(\pi, \pi,0)$ has been found. It is therefore suggested that the BiS$_2$-based superconductors are conventional phonon-mediated superconductors closed to charge density wave instability.\cite{T3,T4} However, this picture was not supported by subsequent neutron scattering experiment.\cite{Lee} The electron-phonon coupling could be much weaker than theoretically expected. It was proposed that the strong Fermi surface nesting would enhance the spin fluctuations, and the electron correlations may play a major role in superconducting pairing.\cite{T6,T2,T8,T9} At present, there is no consensus on the origin of superconductivity in those compounds. It is important to distinguish between different pictures by performing various experimental measurements.

Most of the early experimental investigations were done on polycrystalline samples due to the lack of single crystals. Recently, millimeter-sized single crystals were successfully grown from alkali metal chloride flux in vacuum,\cite{E30,E32} and two angle resolved photoemission spectroscopy (ARPES) studies on the electronic structures of NdO$_{1-x}$F$_x$BiS$_2$ were reported.\cite{E50,E49} Ye et al. found that the electron correlation is very weak and the carrier doping is much smaller than that expected from the nominal fluorine component which they attributed to the Bismuth deficiency.\cite{E50} Another ARPES report by Zeng et al. based on our samples also indicates rather small electron pockets around X point corresponding to an electron doping level of roughly 7\% per Bi site.\cite{E49} Consequently, most of previous theoretical studies based on the high electron doping (close to x=0.5) becomes unreliable, and more experiments based on the single crystals are highly desired.

In this work, we report an optical spectroscopy study on Nd(O,F)BiS$_2$ single crystals in combination with the first principle band structural calculations. The optical measurement reveals a simple metallic response with a plasma frequency of 2.1 eV. Taking the doping level of 7$\%$ of itinerant electrons per Bi site as determined by ARPES experiment on the same batch of crystals, the band structure calculations yield almost the same plasma frequency. Furthermore, the interband transitions at higher energies observed in optical measurement can also be well reproduced by the band structure calculations. Those results reveal essentially an absence of correlation effect in the compound. The study also indicates that the doping level of superconducting sample is far from CDW instability, which was widely discussed for a doping level of x=0.5.

\section{\label{sec:level2}Experiment}

F-doped NdOBiS$_2$ superconducting single crystals were grown using KCl/LiCl (molar ratio KCl:LiCl=3:2) flux method. The raw materials with nominal composition of NdO$_{0.7}$F$_{0.3}$BiS$_2$ were weighted and mixed with the KCl/LiCl flux. The crystal growth procedure is similar to that described by Tanaka\cite{E30} except that we use a different molar ratio of KCl/LiCl:NdO$_{0.7}$F$_{0.3}$BiS$_2$ = 25:1. Many dark-grey shiny plate-like single crystals were obtained. The typical size is 1-2 mm, and obvious layered structure can be observed under the microscope with a typical thickness of $\sim 60 \mu$m.

The obtained single crystals were characterized by X-ray diffraction (XRD) and scanning electron microscope equipped with energy dispersion X-rays spectrum (EDX). Bulk magnetization was measured using a Quantum Design superconducting quantum interference device (SQUID-VSM). Temperature-dependent electrical resistivity was measured by a standard four-probe method in a Quantum Design Physical Property Measurement System (PPMS). The optical reflectance measurements were performed on Bruker IFS 113v and 80v spectrometers in the frequency range from 80 to 24 000 \cm (10 meV$\sim$3 eV). An in situ gold and aluminum overcoating technique was used to obtain the reflectivity R($\omega$). The real part of conductivity $\sigma_1(\omega)$ is obtained by the Kramers-Kronig transformation of R($\omega$). The Hagen-Rubens relation was used for low frequency extrapolation; at high frequency side a $\omega^{-1.2}$ relation was used up to 300 000 \cm, above which $\omega^{-4}$ was applied.

We also performed first-principle calculations for the electronic band structure of NdOBiS$_2$ on the basis of the real crystal structure. The optical constants were calculated and compared with experimental results. A detailed description about the calculations will be presented in the following section.

\section{\label{sec:level2}Results and discussion}

Figure 1 shows the XRD pattern for the Nd(O,F)BiS$_2$ single crystal at room temperature. Only (0 0 $l$)
peaks were observed, indicating a well developed ab-plane orientation for the crystal sheets. For the as-grown crystals, the c-axis lattice parameters was obtained as 13.49 {\AA}, in consistent with the previous reports.\cite{E30,E32}
The averaged composition of the as-grown single crystals is approximately Nd$_{0.95\pm0.02}$O$_y$F$_{0.44\pm0.1}$Bi$_{0.94\pm0.02}$S$_2$ as determined by EDX analysis on several pieces of samples. The composition was normalized to S=2. No Li, K, Cl were detected in the crystals. In the following part, we use nominal component Nd(O,F)BiS$_2$ to represent the actual composition Nd$_{0.95\pm0.02}$O$_y$F$_{0.44\pm0.1}$Bi$_{0.94\pm0.02}$S$_2$.

\begin{figure}[b]
\includegraphics[clip,width=3in]{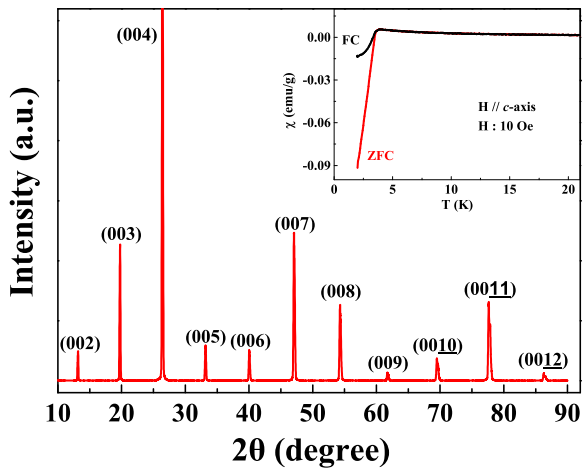}
\caption{(Color online) X-ray diffraction pattern of the Nd(O,F)BiS$_2$ single crystal. Inset shows the temperature dependence of magnetic susceptibility for the annealed Nd(O,F)BiS$_2$ single crystal.}
\end{figure}

Figure 2 shows temperature dependent in-plane resistivity for Nd(O,F)BiS$_2$ single crystals. The curves are normalized to the values at 300 K. For the as-grown crystal, the resistivity values decrease with decreasing the temperature from 300 K, but increase below 100 K before it undergo the superconducting transition with T$_c^{onset}$ = 5.7 K and  T$_c^{zero}$ = 3.6 K. The low temperature upturn is likely due to the disorder/defects of the crystals during the growth procedure. In order to reduce the effect, we have sealed the as-grown crystals in an evacuated silica tube and annealed them at 400 $^\circ \mathrm{C}$ for 10 hours. After such heat treatment, the in-plane resistivity $\rho_{ab}$ reveals a metallic behavior over the entire temperature range between room temperature and T$_c^{zero}$ = 4 K. The zero-field-cooled (ZFC) and field-cooled (FC) dc magnetic susceptibility of the annealed Nd(O,F)BiS$_2$ crystals with $H$ = 10 Oe along c-axis is displayed in the inset of Fig. 1. A sharp diamagnetic signal is observed below 3.8 K, indicating a bulk superconductor. The positive background signal could be due to the paramagnetism of the Nd ion in the crystals. Apparently, the annealing reduces the defects and improves the quality of the crystals. The optical spectroscopy results presented below are based on the annealed Nd(O,F)BiS$_2$ single crystals.

\begin{figure}
\includegraphics[clip,width=3.3in]{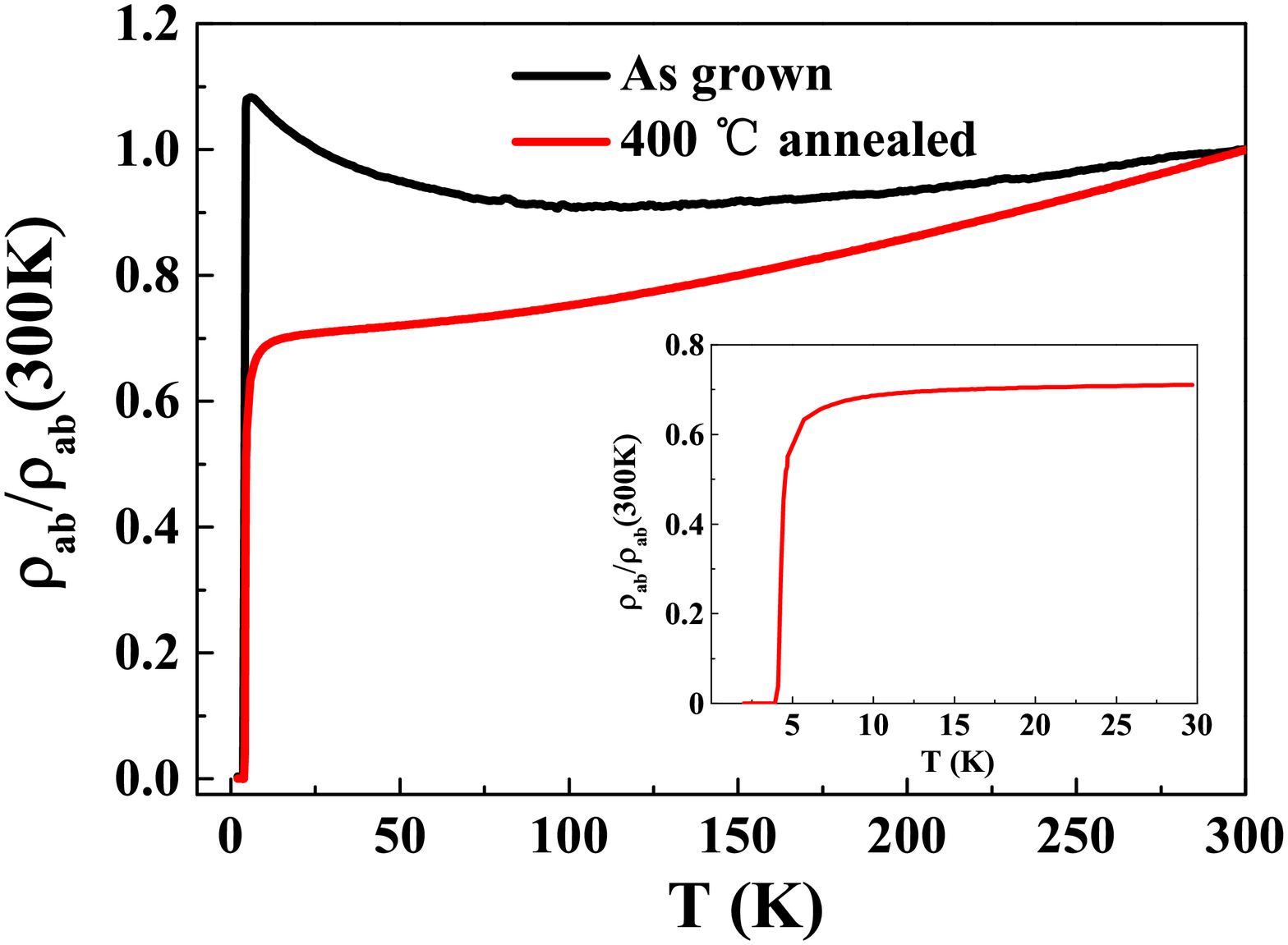}
\caption{(Color online) The normalized electrical resistivity \emph{vs} temperature of both as-grown and annealed Nd(O,F)BiS2 crystals. The inset displays an enlarged region near superconducting transition.}
\end{figure}

Figure 3 (a) and (b) show the reflectance and real part of conductivity spectra of Nd(O,F)BiS$_2$ single crystal over a broad energy scale at different temperatures, respectively. The value of R($\omega$) at low frequency is high and increases further with decreasing temperature, yielding compelling evidence for a good metallic response. With increasing frequency, R($\omega$) drops quickly to a minimum value near 5000 \cm, which is known as the screened plasma edge. Above the edge frequency, the reflectance becomes roughly temperature independent. The relatively low edge position reveals a low carrier density, being consistent with a low electron doping level revealed by ARPES.\cite{E49} As displayed in Fig. 3 (b), the Drude-like conductivity are observed for all
spectra at low frequencies. At high energies, several interband transition features, for example, at 10 500 \cm, 14 400 \cm, and a weak feature near 18 000 \cm, could be well resolved.

\begin{figure}
\includegraphics[clip,width=3.3in]{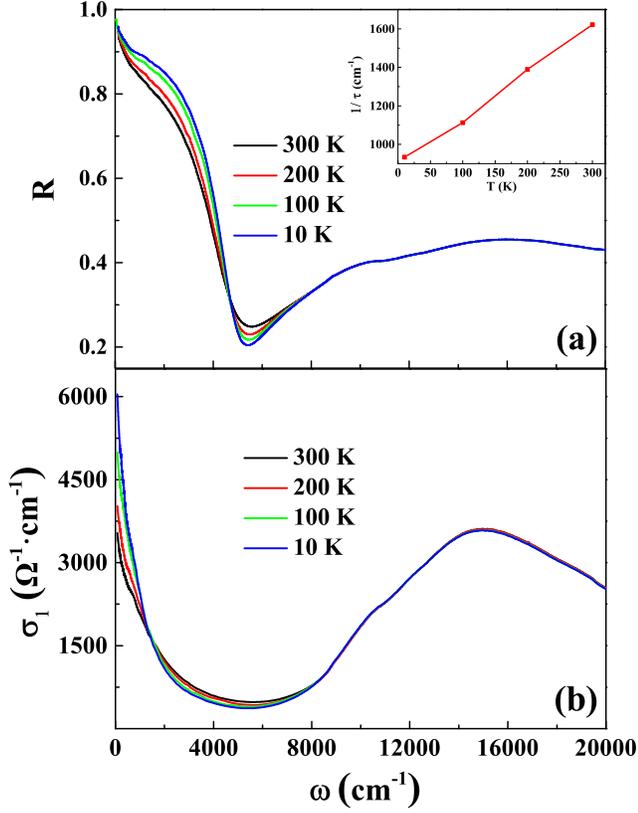}
\caption{(Color online) (a)Temperature dependence of R($\omega$) from 100 to 20000 \cm. (b) Temperature dependence of the real part of the optical conductivity $\sigma_1(\omega)$.}
\end{figure}

To make a quantitative analysis of different contributions to the electronic excitations, we decompose the optical conductivity using a simple Drude-lorentz model:\cite{D-L}
\begin{equation}
\epsilon(\omega)=\epsilon_\infty-{{\omega_{p}^2}\over{\omega^2+i\omega/\tau}}+\sum_{i}{{\Omega_i^2}\over{\omega_i^2-\omega^2-i\omega/\tau_i}}.
\label{chik}
\end{equation}
where $\epsilon_\infty$ is the dielectric constant at high energy, and the middle and last terms are the Drude and Lorentz components, respectively. The Drude component represents the contribution from conduction electrons, while the Lorentz components describe the interband transitions. The optical conductivity spectra below 20 000 \cm could be reasonably fit by one Drude and three Lorentz components. As an example, we show in Fig. 4 the room temperature spectrum and decomposed Drude and Lorentz components. The plasma frequency obtained for the Drude components is 17 000 \cm($\sim$ 2.1 eV), and it keeps roughly unchange at different temperatures, while the width, i.e. the scattering rate 1/$\tau$, in the Drude term decreases with decreasing temperature, as displayed in the inset of Fig. 3 (a). The three Lorentz terms centered at 10 500, 14 400 and 17 800 \cm, respectively, are attributed to the interband transitions, and their origins will be discussed below.

\begin{figure}
\includegraphics[clip,width=3.3in]{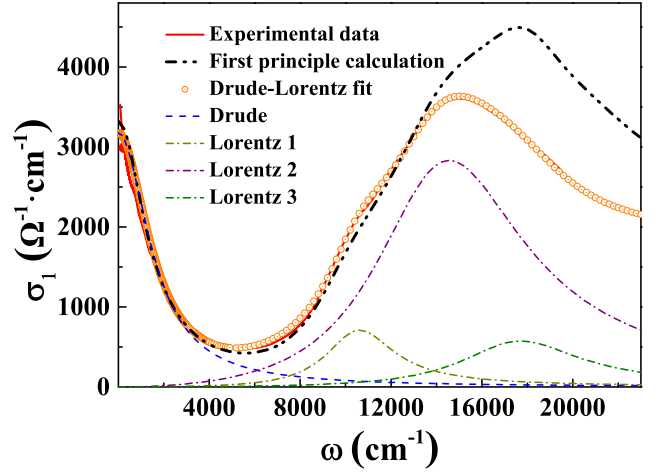}
\caption{(Color online) The real part of the optical conductivity $\sigma_1(\omega)$ at 300 K, together with the decomposed Drude and Lorentz components. The calculated conductivity based on the first-principle band structure within GGA and an assumed value of scattering rate is also presented.}
\end{figure}

According to band structural calculations, the Fermi surfaces are big for the doping level of x=0.5, consisting of large pockets encircling the center $\Gamma$ and and corner M of the Brillouin zone (BZ), respectively. Correspondingly, the calculated plasma frequency for LaO$_{0.5}$F$_{0.5}$BiS$_2$ is 5.99 eV.\cite{T3} This value is much larger than the value determined by optics in this work for Nd(O,F)BiS$_2$ single crystal. However, as we explained in the introduction, the recent ARPES measurements on Nd(O,F)BiS$_2$ actually revealed only smaller Fermi pockets encircling the midpoints on the edge of BZ with a doping level of roughly 7$\%$ per Bi site.\cite{E49} On this account, we have to compare the experimental results to theoretical calculations only at such small doping level.

We performed the first-principle calculations of electronic structure by using the full-potential linearized-augmented plane-wave(FP-LAPW) method implemented in WIEN2K package for the real crystal structure of undoped NdOBiS$_2$\cite{WIEN2k}. The exchange-correlation potential was treated using the generalized gradient approximation(GGA) based on the Perdew-Burke-Ernzerhof(PBE) approach\cite{PBE}. Spin-orbit coupling(SOC) was included as a second variational step self-consistently. The radii of the muffin-tin sphere $R_{MT}$ were 2.37 Bohr for Nd, Bi and S and 2.1 Bohr for O, respectively. A 18$\times$18$\times$6 k-point mesh has been utilized in the self-consistent calculations. For the optical properties calculation\cite{optics} a fine grid mesh with 25$\times$25$\times$7 was adopted. The truncation of the modulus of the reciprocal lattice vector $K_{max}$, which was used for the expansion of the wave functions in the interstitial region, was set to $R_{MT}*K_{max}=7$. This parent compound is a band semiconductor and its Fermi level (E$_F$, set as zero energy) locates in the energy gap. With electron doping by F substitution for O, the Fermi level should shift up. Figure 5 (a) and (b) shows the calculated band dispersions and the Fermi surfaces with chemical potential being shifted up by 0.85 eV, which roughly corresponding to a doping level of 7$\%$ per Bi site. The shape and Fermi surface areas are in good agreement with the ARPES measurement on the same batch of crystals.\cite{E49} The calculated plasma frequency at this chemical potential is 2.078 eV, which is very close to the experimental value. Figure 5 (c) shows the real and imagine parts of dielectric function contributed for the interband transitions below 4 eV, the free carrier contributions are not considered in this plot.

\begin{figure}
\includegraphics[clip,width=3.3in]{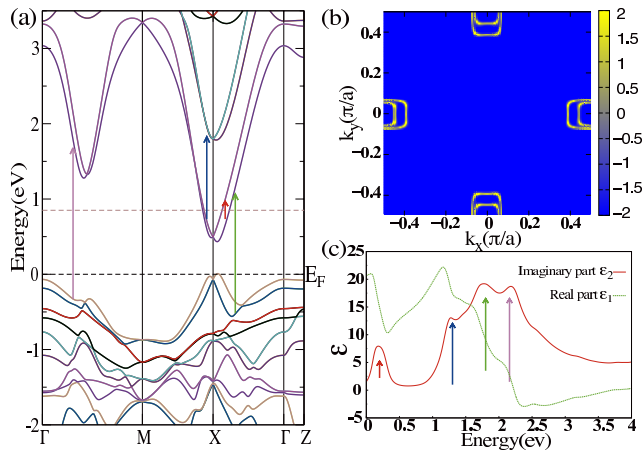}
\caption{(Color online) (a) The band dispersions of NdOBiS$_2$ by first-principle calculations within GGA with a shift of E$_F$ by 0.85 eV. (b) The Fermi surfaces of NdOBiS$_2$ obtained by shifting up E$_F$$\sim$0.85 eV, which roughly corresponding to a doping level of 7$\%$ per Bi site. (c) the real part and imaginary part of dielectric function. The interband transition peaks and corresponding transitions in the band dispersions are indicated by the arrows.}
\end{figure}

Since the scattering rate of free carriers is not a theoretically calculated quantity, the spectral shape of Drude component could not be given by first-principle band calculations. For the reason of making a comparison with experiment, we add a value of the scattering rate for the Drude component by hand. Then the spectrum of the Drude part could also be displayed. The calculated real part of conductivity by assuming a scattering rate value of 1600 \cm is also displayed Fig. 4. Surprisingly, we find that the calculated conductivity curve can well reproduce the experimental curve not only for the low frequency Drude component but also for the energy scales of the interband transitions. The different values of the conductivities at high energy higher than 15 000 \cm could be largely attributed to the high energy extrapolations in Kramers-Kronig transformation.

According to the band structural calculations, there should be four major interband transitions below the energy of 4 eV, as indicated by the arrows in Fig. 5. Those interband transitions lead to peaks in the imaginary part of the dielectric function. The peak at the lowest energy near 0.25 eV is from the transition between two parallel bands from Bi p orbitals being caused by the Bi-Bi inter-layer coupling. The peak near 1.25 eV is from the transition between two parallel bands from Bi p$_x$ and p$_y$ orbitals arising from the in-plane p-p $\sigma$-bond and p-p $\pi$-bond. The peaks near 1.8 eV and 2.1 eV both come from the transition between S or O p-bands to Bi p-bands. The last three could be well resolved in the optical conductivity spectra at respective energies. However, the lowest interband transition near 0.25 eV is embedded in the strong peak from the Drude response of free carriers and could hardly be distinguished from the experimental data. Since the plasma frequency determined by experiment is even slightly higher than that of first-principle calculations, it is very likely that such interband transition is indeed present in the experimental data. Nevertheless, this interband transition must be extremely weak in intensity and only a very small fraction of spectral weight is taken from the Drude component.

Usually, the ratio of experimental kinetic energy K$_{expt}$ to that of the theoretical kinetic energy K$_{band}$ from density functional calculations provides a measure of the degree of correlation.\cite{Basov,Qimiao} The kinetic energy is proportional to the spectral weight $\omega_{p}^2$/8 defined as the area under the Drude part of $\sigma_1(\omega)$. The extremely close values of the plasma frequencies determined by both experiment and GGA calculations implies $K_{expt}/K_{band}=\omega_{p,expt}^2/\omega_{p,band}^2 \approx 1$, suggesting that the Nd(O,F)BiS$_2$ crystal is a simple metal and correlation effect is almost absent. The result is consistent with the recent ARPES report.\cite{E50}

The absence of the electronic correlation effect and the surprisingly good agreement of optical data between the experimental measurement and the first-principle calculations indicate that the BiS$_2$-based compound is an ordinary metal. The scenario for the superconductivity based on the exotic pairing mechanism or strong electronic correlation effect is not well grounded. The small plasma frequency, which is in good agreement with the small Fermi surfaces observed from ARPES measurements, suggests that the compound is far from the CDW instability at the doping level of x=0.5 as proposed in early theoretical calculations. As proposed in early work,\cite{E50} the smaller doping level could be due to the presence of Bi vacancies in the samples.

\section{\label{sec:level2}Summary}

To summarize, we have grown F-substituted NdOBiS$_2$ superconducting single crystals using KCl/LiCl flux method. The annealed crystals show a typical metallic behavior with a superconducting transition temperature near 4 K. We performed a combined optical spectroscopy and first principle band structural calculation study on Nd(O,F)BiS$_2$ single crystal. The optical measurement reveals a simple metallic behavior with a relatively small plasma frequency of 2.1 eV. Taking the doping level of 7$\%$ of itinerant electrons per Bi site as determined by ARPES experiment on the same batch of crystals, the band structure calculations not only well reproduce the spectral weight of the low frequency Drude component but also the energy scales of the interband transitions. Those results indicate essentially the absence of correlation effect in the compound. The study also illustrates that the doping level of superconducting compound is far from the CDW instability suggested in early studies for a doping level of x=0.5.

\begin{acknowledgments}

This work is supported by the National Science Foundation of
China (11120101003, 11327806), and the 973 project of the
Ministry of Science and Technology of China (2011CB921701, 2012CB821403).

\end{acknowledgments}


\begin{references}
\bibitem{E1} Yoshikazu Mizuguchi, Hiroshi Fujihisa, Yoshito Gotoh, Katsuhiro Suzuki, Hidetomo Usui, Kazuhiko Kuroki, Satoshi Demura, Yoshihiko Takano, Hiroki Izawa, Osuke Miura, Phys. Rev. B \textbf{86}, 220520(R) (2012).
\bibitem{E6} Shiva Kumar Singh, Anuj Kumar, Bhasker Gahtori, Shruti Kirtan, G. Sharma, S. Patnaik, V. P. S. Awana, J.Am. Chem. Soc., \textbf{134}, 16504 (2012).
\bibitem{E2} Yoshikazu Mizuguchi, Satoshi Demura, Keita Deguchi, Yoshihiko Takano, Hiroshi Fujihisa, Yoshito Gotoh, Hiroki Izawa, Osuke Miura, J. Phys. Soc. Jpn. \textbf{81}, 114725 (2012).
\bibitem{E4} Satoshi Demura, Yoshikazu Mizuguchi, Keita Deguchi, Hiroyuki Okazaki, Hiroshi Hara, Tohru Watanabe, Saleem James Denholme, Masaya Fujioka, Toshinori Ozaki, Hiroshi Fujihisa, Yoshito Gotoh, Osuke Miura, Takahide Yamaguchi, Hiroyuki Takeya, Yoshihiko Takano, J. Phys. Soc. Jpn. \textbf{82}, 033708 (2013).
\bibitem{E13} Jie Xing, Sheng Li, Xiaxing Ding, Huang Yang, Hai-Hu Wen, Phys. Rev. B \textbf{86}, 214518 (2012).
\bibitem{E15} Rajveer Jha, Anuj Kumar, Shiva Kumar Singh, V. P. S. Awana, J. Sup. and Novel Mag. \textbf{26}, 499-502 (2013).
\bibitem{E16} D. Yazici, K. Huang, B. D. White, A. H. Chang, A. J. Friedman, M. B. Maple, Philosophical Magazine \textbf{93}, 673 (2012).
\bibitem{E19} Xi Lin, Xinxin Ni, Bin Chen, Xiaofeng Xu, Xuxin Yang, Jianhui Dai, Yuke Li, Xiaojun Yang, Yongkang Luo, Qian Tao, Guanghan Cao, Zhuan Xu, Phys. Rev. B \textbf{87}, 020504(R) (2013).
\bibitem{E21} D. Yazici, K. Huang, B. D. White, I. Jeon, V. W. Burnett, A. J. Friedman, I. K. Lum, M. Nallaiyan, S. Spagna, M. B. Maple, Phys. Rev. B \textbf{87}, 174512 (2013).
\bibitem{E36} A.Krzton-Maziopa, Z. Guguchia, E. Pomjakushina, V. Pomjakushin, R. Khasanov, H. Luetkens, P. Biswas, A. Amato, H. Keller, K. Conder,  J. Phys.: Condens. Matter \textbf{26}, 215702 (2014).
\bibitem{cuprates} Warren E. Pickett, Rev. Mod. Phys. \textbf{61}, 433 (1989).
\bibitem{iron-based}  Y.Kamihara, T.Watanabe, M.Hirano, and H.Hosono, J.Am. Chem. Soc., \textbf{103}, 3296(2008).
\bibitem{T1} Hidetomo Usui, Katsuhiro Suzuki, and Kazuhiko Kuroki, Phys. Rev. B \textbf{86}, 220501(R) (2012).
\bibitem{T3} Xiangang Wan, Hang-Chen Ding, Sergey Y. Savrasov, Chun-Gang Duan, Phys. Rev. B \textbf{87}, 115124 (2013).
\bibitem{T4} B. Li, Z. W. Xing, G. Q. Huang, Europhys. Lett. \textbf{101}, 47002 (2013).


\bibitem{Lee} J. Lee, M. B. Stone, A. Huq, T. Yildirim, G. Ehlers, Y. Mizuguchi, O. Miura, Y. Takano, K. Deguchi, S. Demura, and S.-H. Lee, Phys. Rev. B \textbf{87}, 205134 (2013).


\bibitem{T6} Taner Yildirim,  Phys. Rev. B \textbf{87}, 020506(R) (2013).

\bibitem{T2} Tao Zhou, Z. D. Wang, Journal of Superconductivity and Novel Magnetism \textbf{26}, 2735 (2013).

\bibitem{T8} Yi Liang, Xianxin Wu, Wei-Feng Tsai, Jiangping Hu, Frontiers of Physics \textbf{9}, 194 (2014).

\bibitem{T9} Y. Yang, W. S. Wang, Y. Y. Xiang, Z. Z. Li, and Q. H. Wang, Phys. Rev. B \textbf{88}, 094519 (2013).

\bibitem{E30} Masanori Nagao, Satoshi Demura, Keita Deguchi, Akira Miura, Satoshi Watauchi, Takahiro Takei, Yoshihiko Takano, Nobuhiro Kumada, Isao Tanaka, J. Phys. Soc. Jpn. \textbf{82}, 113701 (2013).

\bibitem{E32} Jianzhong Liu, Delong Fang, Zhenyu Wang, Jie Xing, Zengyi Du, Xiyu Zhu, Huan Yang, Hai-Hu Wen,  Europhys. Lett. \textbf{106}, 67002 (2014).
.
\bibitem{E50} Z. R. Ye, H. F. Yang, D. W. Shen, J. Jiang, X. H. Niu, D. L. Feng, Y. P. Du, X. G. Wan, J. Z. Liu, X. Y. Zhu, H. H. Wen, M. H. Jiang, e-print arXiv:1402.2860.

\bibitem{E49} L. K. Zeng, X. B. Wang, J. Ma, P. Richard, S. M. Nie, H. M. Weng, N. L. Wang, Z. Wang, T. Qian, H. Ding, e-print arXiv:1402.1833.

\bibitem{D-L} W. Z. Hu, J. Dong, G. Li, Z. Li, P. Zheng, G. F. Chen, J. L. Luo, and N. L. Wang, Phys. Rev. Lett. \textbf{101}, 257005 (2008).


\bibitem{WIEN2k} P. Blaha, K. Schwarz, G. Madsen, D. Kvaniscka, and J. Luitz, Wien2k, An Augmented Plane Wave Plus Local Orbitals Program for Calculating Crystal Properties (Vienna University of Technology, Vienna,Austria, 2001).

\bibitem{PBE} J. P. Perdew, K. Burke and M. Ernzerhof, Phys. Rev. Lett. \textbf{77}, 3865 (1996).

\bibitem{optics} R. Abt, C. Ambrosch-Draxl and P. Knoll, Physica B \textbf{194-196}, 1451 (1994); C. Ambrosch-Draxl and J. O. Sofo, Comput. Phys. Commun. \textbf{175}, 1 (2006).


\bibitem{Basov} M. M. Qazilbash, J. J. Hamlin, R. E. Baumbach, Lijun Zhang, D. J. Singh, M. B. Maple, D. N. Basov, Nature physics \textbf{5}, 647 (2009).

\bibitem{Qimiao} Qimiao Si, Nature physics \textbf{5}, 629 (2009).




\end{references}
\end{document}